\documentstyle[a4]{article}

\begin{document}

\newcommand{\bib}{\bibitem}
\newcommand{\er}{\end{eqnarray}}
\newcommand{\br}{\begin{eqnarray}}
\newcommand{\be}{\begin{equation}}
\newcommand{\ee}{\end{equation}}
\newcommand{\epe}{\end{equation}}
\newcommand{\bea}{\begin{eqnarray}}
\newcommand{\eea}{\end{eqnarray}}
\newcommand{\ba}{\begin{eqnarray}}
\newcommand{\ea}{\end{eqnarray}}
\newcommand{\epa}{\end{eqnarray}}
\newcommand{\ar}{\rightarrow}
\def\I{{\cal I}}
\def\A{{\cal A}}
\def\F{{\cal F}}
\def\a{\alpha}
\def\b{\beta}
\def\r{\rho}
\def\D{\Delta}
\def\R{I\!\!R}
\def\l{\lambda}
\def\d{\delta}
\def\T{\tilde{T}}
\def\k{\kappa}
\def\t{\tau}
\def\f{\phi}
\def\p{\psi}
\def\z{\zeta}
\def\G{\Gamma}
\def\ep{\epsilon}
\def\hx{\widehat{\xi}}
\def\na{\nabla}
\newcommand{\bslash}{b\!\!\!/}
\newcommand{\vslash}{v\!\!\!/}
\newcommand{\eslash}{e\!\!\!/}
\newcommand{\rslash}{r\!\!\!/}


\begin{center}

{\large Doublet Groups, Extended Lie Algebras, and Well Defined
Gauge Theories for the Two Form Field.}

\vspace{0.8cm}

 Marcelo Botta Cantcheff\footnote{botta@cbpf.br, mbotta$\_$c@ictp.trieste.it}

\vspace{4mm} High Energy Section,

 Abdus Salam ICTP, Strada Costiera 11, 34014 Trieste, Italy.

\vspace{3mm}

 Centro Brasileiro de Pesquisas Fisicas (CBPF)

Departamento de Teoria de Campos e Particulas (DCP)

Rua Dr. Xavier Sigaud, 150 - Urca

22290-180 - Rio de Janeiro - RJ - Brazil\footnote{Permanent
Address.}.
\end{center}

\begin{abstract}

 We propose a symmetry law for a doublet of different form
fields, which resembles gauge transformations for matter fields.
This may be done for general Lie groups, resulting in an extension
of Lie algebras and group manifolds. It is also shown that
non-associative  algebras naturally appear in this formalism,
which are briefly discussed.

Afterwards, a general connection which includes a two-form field
is settled-down, solving the problem of setting a gauge theory for
the Kalb-Ramond field for generical groups.

Topological Chern-Simons theories can also be defined in four
dimensions, and this approach clarifies their relation to the
so-called $B \wedge F$-theories. We also revise some standard
aspects of Kalb-Ramond theories in view of these new perspectives.

Since this gauge connection is built upon a pair of fields
consisting of a one-form and a two-form, one may define Yang-Mills
theories as usually and, remarkably, also {\it minimal coupling}
with bosonic matter, where the Kalb-Ramond field appears naturally
as mediator; so, a new associated conserved charge can be defined.
For the Abelian case, we explicitly construct the minimal
interaction between $B$-field and matter following a "gauge
principle" and find a novel conserved tensor current. This is our
most significative result from the physical viewpoint.

 This framework is also generalized in such a way that any $p$-rank tensor
  may be formulated as a gauge field.

\end{abstract}

\section{Introduction}

The (Abelian) Kalb-Ramond field \cite{kr0,kr} (KR), $B_{\mu\nu}$,
is a two-form field which appears in the low energy limit of
String Theory \cite{it5}, in Quantum Gravity \cite{it6} and in
several other frameworks in Particle Physics \cite{aplic}. In
particular, most attempts to incorporate mass to gauge field
models in four dimensions take into account this object added to a
one form gauge field \cite{tm0,tm,la}. However, their actual
underneath group structure is lacking. It is often implemented
{\it by hand} in order to analyze the gauge invariance of certain
$B\wedge F$ models.

The symmetry of the KR field is remarkably similar to that of a
$1$-form gauge field \cite{la}:
\begin{equation}
B_{\mu\nu} \to B_{\mu\nu} + \partial_{[\mu}\beta_{\nu]},
\label{trKR}
\end{equation}
where $\beta_\nu$ is a $1$-form parameter. The question is: how
can we associate the parameter $\beta_{\mu}$ to the manifold of
some gauge group \cite{otro,ultimo}? This problem was rigorously
analyzed in refs. \cite {gkr,krvec} where the representations were
singlet tensor/spinor spaces with inner product, and the KR field
was built in the connection \cite {dkr,krvec}. However, many
difficulties arose involving Lorentz invariance of physical models
in the non-Abelian case \cite{dkr,krvec,teit}. Also, it was not
clear how the KR field could be built when spacetime would be
non-flat. In this paper we propose a framework where these
difficulties are solved.

From the physical point of view, it is essential to ask if a {\it
genuine} gauge theory may be formulated for this field, {\it i.e},
if the two-form gauge potential may be stated as a connection on
some group manifold. This is important because, as it is known,
this structure would be crucial for the identities which determine
the finiteness or not of physical models. In particular, in ref.
\cite{h}, it was proven that massive (non-Abelian) gauge models
\cite{tm0,tm} necessarily based on a gauge KR field, $B_{\mu\nu}$,
and a usual one-form, $A_{\mu}$, are ill defined in four spacetime
dimensions. The first objective of this article is to establish
clearly this group structure and to show that these theories may
be formulated in a similar way as the Yang-Mills-Chern-Simons
theories in $2+1$ dimensions, which are known to be finite
\cite{helpi}.

Another crucial question is how to define a minimal coupling of
this field with matter fields, with the interaction with gauge
fields appearing by replacing partial derivatives of the matter
fields by covariant ones in the free Lagrangian. This is directly
related to the charge conservation laws via Noether's theorem. To
do this, local gauge transformations for matter fields need to be
defined. Up to now, this is unknown for transformations which
involve a $1$-form parameter. This is our second objective here.
Recently, other perspectives have been presented for these
questions \cite{0it}, where its expected applications in
gravitation with torsion and Kalb-Ramond cosmology are mentioned
\cite{0it,it11,it12,it13}.

To tackle all these problems, we explore here a new possibility in
order to have a well defined two form gauge field: relaxing the
requirement of singlet tensorial representations\footnote{Namely,
a group representation given by a single tensor or spinor (or
tensor product of them).}, imposed in preceding approaches
\cite{gkr,dkr,krvec}. We shall show that this allows us to
construct well defined gauge models for KR fields, which may be
minimally coupled with matter fields in a natural way. Once more,
the simplest solution of the problem arises from considering
doublets of tensors of different ranks as a representation for a
Lie group. This kind of idea has been successfully used to solve
other algebraic questions related to Hodge duality \cite{dob}.

By considering a doublet field representation, we are able to
include an $1$-form parameter in an {\it exponential-like}
symmetry/transformation law: \be\label{transdob0} \d \left(
\begin{array}{c}
  \phi \\
 \phi_\mu
\end{array}\right)= \left( \begin{array}{c}
i\alpha \phi + i\b^\mu \phi_\mu\\ i \alpha \phi_\mu + i \b_\mu
\phi
\end{array}\right)
= i \left( \begin{array}{cc}
  \a & \b \\
  \b & \a
\end{array}\right)
\left(\begin{array}{c}
  \phi \\
  \phi_\mu
\end{array}\right)
\ee where the variation of the fields is proportional to
themselves and to the group parameters. These simple expressions
solve the problem of writing this kind of transformation law in a
simple and satisfactory way, and are the key to define the group
operations involving an $1$-form parameter. Notice that, without a
doublet representation (and a scalar parameter $\alpha$),
individual fields ($\phi$ and $\phi_\mu$) can never be combined
with an $1$-form $\beta_\mu$ to give a tensor of the same type,
and to define their variations. Furthermore, in ref. \cite{krvec}
 (also in \cite{teit})
it has been shown that, if one insists in representing such groups
with single fields, then $\beta$ must be decomposed with respect
to an orthogonal spacetime basis and the Lorentz symmetry is
broken in gauge theories, except in the Abelian case, where
relativistic invariance is restored in the gauge actions. In such
context, it is also unclear how can it be generalized to non-flat
base manifolds .

 Clearly, (\ref{transdob0}) is the most general rule
 where both $\b$ and a minimal number of matter
 fields appear linearly and in a Lorentz-invariant way, such as
 was argued in refs. \cite{krvec,teit}.
  This idea may sound technically trivial but it is meaningful, it has never been used
  before as the cornerstone for
 a gauge principle generating the two-form field.

This approach is more satisfactory than previous ones
\cite{dkr,krvec}, as everything can be expressed in a manifestly
covariant form, {\it i.e.}, we do not need to define the
representation with respect to a spacetime coordinate basis, and
the generalization to curved spacetime becomes rather immediate.

This work is organized as follows: in Section 2, we explicitly
find out the Lie group corresponding to these transformations and
the covariant derivative with the generic tensor field being part
of the connection is defined, and in Section 3, gauge theories as
Yang-Mills and Chern-Simons in four dimensions are discussed.
Finally, in Section 4, we study the minimal interaction of this
doublet connection with matter. Concluding remarks are collected
in Section 5.

\section{Doublet Field Representations and
 Tensorial \\ Symmetry  Parameters.}

Let $(M, g_{\mu \nu})$ be a four-dimensional oriented spacetime
and $G$ be a Lie group whose associated algebra is ${\cal G}$;
${\tau}^{a}$ are the matrices representing the generators of the
group with $a= 1,\ldots , \mbox{dim}\:G$; $\tau_{abc}$ are the
structure constants.

As mentioned, consider the general transformations
\bea  \d\phi =  i\alpha \phi + i \b^\mu \phi_\mu \nonumber\\
\d\phi_\mu =i \alpha \phi_\mu + i \b_\mu \phi \label{transdob}\eea
where the doublet of parameters $(\alpha , \beta)$ consists of two
Lie algebra valued $0$- and $1$-forms respectively \footnote{We
assume them in a matricial representation of the algebra.}. Let us
denote the doublet of fields by $\Phi \equiv (\phi ,\phi _{\mu
})$. Thus, this transformation may be formally expressed as \be
\delta_{\a, \b} \Phi = ( I\a + \sigma \b ) \Phi \ee where
$I,\sigma $ are $2\times 2$ matrices, the identity and the first
of Pauli's matrices (often denoted by $\sigma _{1}$) respectively.
The product of two elements of the algebra is well defined and is
naturally given by the usual matrix product \be \label{matprod}
\delta_{\a', \b'} \,\delta_{\a, \b} \Phi =  I (\a'\, \a + b
\b'_\mu \, \b^\mu) + \sigma (\a \,\b' + \a' \,\b) \Phi , \ee where
we have introduced the real parameter $b$ multiplying the metric
$g_{\mu \nu}$, a priori taken to be equal to one. However, one may
explicitly verify that this algebra is non-associative, due
precisely to the term which is quadratic in $\beta $, since it
involves scalar products. In structures like these (called {\it
quasi-algebras}) the Jacobi identity must be replaced by a weaker
expression \cite{asoc}. This quasi-algebra generates a
quasi-group, which has all properties of a group, except
associativity. In the present case, associativity is satisfied for
subsets of $\beta $-parameters which are all {\it parallel}
between themselves. So, the set of Lie parameters may be thought
to describe a collection of groups (parameterized by the set of
orientations in spacetime). This coincides remarkably with the
main result of ref. \cite{gkr}, where singlet representations are
analyzed.

Despite this, we can repeat most of the steps towards well defined
theories with non-associative gauge symmetry. For instance, we
could formally define a covariant derivative and a curvature
tensor since, only the infinitesimal structure is required.
 We will return to this point in Sub-Section 2.1.

The whole structure may alternatively be expressed in terms of
doublets, ordered pairs of $0$- and $1$-forms. The product of
doublets reads as: \be \left(  \alpha\;,\;\beta_{\mu}\right)
\left(  \phi\;,\;\phi_{\mu
}\right)  =\left(  \phi\alpha+b\phi_{\mu}\beta^{\mu}\;,\;\phi_{\mu}%
\alpha+\phi\beta_{\mu}\right) \label{defprod} \ee thus
\[
\left(  \delta\phi\;,\;\delta\phi_{\mu}\right)  =i\left(
\alpha\;,\;\beta_{\mu}\right)  \left(  \phi\;,\;\phi_{\mu}\right)
\]

Besides non-associativity, a second problem with this product is
that, since it involves a spacetime metric, it cannot be a
topological construction. However, this does not constitute a
technical difficult in itself, if one is not interested in
topological theories.

In order not to deal, in this paper, with the two problems
mentioned before (which would conduct us to many interesting
possibilities), we will concentrate in a $b$-product with $b=0$,
which is associative and defines a Lie algebra.
In this case, the symmetry transformation reads
  \be  \left( \begin{array}{c}
 \d \phi \\
 \d \phi_\mu
\end{array}\right)= \left( \begin{array}{c}
 i\phi \alpha \\ i \phi_\mu \alpha + i \phi \b_\mu
\end{array}\right) .\ee

For simplicity, we will consider a group structure $G = G_{(\a)}
\times G_{(\b)}$, where $\a=\a^a \tau^a \in {\cal G}_{(\a)} \; ,
\b=\b_\mu^a \tau'^a \in {\cal G}_{(\b)}$ ($\b \in {\cal
G}_{(\b)}\otimes \Lambda_1$) are Lie algebra valued. Clearly, $[\a
,\b]=0$ . In this case, the calculation of explicit form for the
group elements  simplifies considerably (see below), however,
there is no technical obstacle in generalizing the procedure to
any Lie algebra (namely, any $G$ where $[\a,\b] \neq 0 $).

The identity for this product is $(1 , 0)$. The following formula
may be easily shown by induction \be\label{pot} (\a , \b)^n =
(\a^n \; , \;n \b_\mu \a^{n-1} ) . \ee Defining $(\ep , \ep_\mu )
= (\frac{\a}{n} , \frac{\b_\mu}{n} )$, an infinitesimal group
transformation can be written \be \Phi'= g(\ep , \ep_\mu ) \Phi =
\Phi + i(\ep , \ep_\mu )\Phi = ( (1,0) + i(\ep , \ep_\mu ) )\Phi =
(1 + i\ep , i\ep_\mu )\Phi .\ee We now compose this operation
$n$-times, according to (\ref{pot}):\bea g\left(
\epsilon,\epsilon_{\mu}\right)  ^{n}\Phi & =\left(
1+i\epsilon,i\epsilon_{\mu}\right)  ^{n}\Phi=\left( \left(
1+i\epsilon\right)  ^{n}, in\epsilon_{\mu}\left(
1+i\epsilon\right)
^{n-1}\right)  \Phi \nonumber \\
& =\left(  \left(  1+i\frac{\alpha}{n}\right)  ^{n},in\frac{\beta_{\mu}}%
{n}\left(  1+i\frac{\alpha}{n}\right)  ^{n-1}\right)  \Phi .\eea
Taking the limit $n\to \infty$, when $n\sim n-1$, we obtain: \be
\lim_{n\to\infty} g(\ep , \ep_\mu )^n  = ( \exp{i\a} , i\b_\mu
\exp{i\a}).\ee Thus, the final closed form for a generic group
element is \be\label{sepg} g(\a , \b_\mu ) = ( e^{i\a} , i \b_\mu
e^{i\a}),\ee which is one of our main results. The inverse element
is $[g(\a , \b_\mu )]^{-1} = g(-\a \, ,\, -\b_\mu )$.

In the Abelian case this has the properties of an exponential
function since \be g(\a , \b_\mu ) g(\a' , \b'_\mu ) = g(\a +\a' ,
\b_\mu + \b'_\mu) \ee and $g(0 , 0)= Id$ $(= (1,0))$. Notice that
in this (separate) case, by virtue of the product defined above,
all group elements may be factorized as: \be g(\a , \b_\mu ) =g(\a
, 0)\, g(0 , \b_\mu )= g(0 , \b_\mu) g(\a , 0 ) .\ee  These
transformations are the crucial point in this paper; after that,
the rest of the construction follows in a straightforward way.

 Let us remark once more that here, for simplicity, we are
going to construct gauge theories for these separable groups,
whose elements are generically expressed by (\ref{sepg}), since
our main objective in this article is to show, in a concise way,
some remarkable theoretical consequences of this formalism (say,
minimal coupling from a gauge principle and the existence of a
vector Noether charge), at least in the simplest situation.
However, for completeness, in the next subsections we briefly
discuss  the other interesting generalizations, including the
non-associative case.

\subsection{Non-Associative Symmetry and Non-Separable Groups.}

In the non-associative case (for instance, with $b=1$), since all
infinitesimal parameters $\ep_\mu$ are considered parallel (the
product is clearly associative in this subset), $g(\a, \b)$ can be
found by a similar procedure: \be \label{gb1} g_{(b=1)}(\a\, ,\,
\b_\mu \equiv e_\mu \b ) \equiv \left(g \left(\frac{\a}{n}\, ,\,
\frac{\b_\mu}{n}\right) \right)^n = (\cos{\b} \, e^{i\a} \, , \, i
e_\mu sin{\b} \, e^{i\a})\; , \ee where $e_\mu$ is a unit
one-form. From this, we can verify directly that, although the
product of these objects (quasi-group elements) is indeed
non-associative, a weaker associativity (quasi-associativity) of
the form: $a(ba) = (a b) a$, is satisfied\footnote{This is
actually stronger: $g_3 (g_2 g_1)= (g_3 g_2 ) g_1 $ if their
corresponding $e_3 , e_1$ coincide.}.

In fact, by composing a large number of infinitesimal, general
transformations (\ref{transdob}), one can represent the result as
an exponential:
 \be\label{exprep} g=\exp{i( \a I + \b \sigma_e )} \ee whose precise meaning is given
 according to the algebra: \be
\sigma_e \sigma_{e'} \,=\, b\, ( e . e') \; I \; ,\ee where $e,e'$
are the respective unit directions of two arbitrary one-form
parameters \footnote{So, these objects may be represented as a
$2\times 2$ matrix, $\sigma_e = e \sigma $ .} . Thus, in the
separate case, this can be expressed as a doublet, (\ref{gb1}).

As it was previously commented, one can construct the same objects
as those which appear in a standard associative gauge theory
(namely, connection, curvature, actions) since this is a gauge
symmetry in its own right. because the associative behavior is
recovered to first order in the parameters.

The exponential notation is convenient even for the associative
case: one can perform calculations and, at the end, take $b\to 0$
to recover a Lie structure. Equivalently, one can consider the
 leading order in $\b$, as we can see directly from
expression (\ref{gb1}). In this case, the algebra above becomes
degenerate and all $ \sigma_e $ may be
 identified with a single $\sigma$ (such that $\sigma^2=0$ ).
So, we can say that in general this is a quasi-group manifold,
which locally (in a neighborhood of $\b=0$), approaches a Lie
group.  In particular, this is the best way to represent an
element of a non-separable group. By truncating (\ref{exprep}) to
first order in $\b$, in the doublet notation, we obtain the
general form of a generic element of an extended Lie group:
\be\label{nosepg}
 g(\a,\b) = \left( e^{i\a}  \, , \,
   \sum_{n=0}^\infty \frac{i^n}{n!} \left[\sum_{m=1}^n \a^{m-1} \b_\mu  \a^{n-m}\right]
   \right),
\ee where $\a , \b_\mu$ are valued in any given Lie algebra ${\cal
G}$. If $[\a,\b]=0$, (\ref{nosepg}) yields (\ref{sepg}), as
expected. Notice that the expression (\ref{nosepg}) is not
convenient for calculations. So, as mentioned, the exponential
notation  (up to $o^2(\b)$ or $o(b)$ contributions) should be used
instead of this.

\subsection{Generalized Doublets and $(0,r)$-Tensors as Gauge
Fields.}

Let us consider general representations of {\it $p-$doublets of
order $r$ }, which consist in pairs $(\phi_p , \phi_{p+r})\in
\Pi_p \times \Pi_{p+r}$, where
 $\Pi_p $ denotes the standard set of tensors of type $(0,p)$.
So, the symmetry transformation can be built over doublets of
order $r$ using the same idea, in any of the spaces $\Pi_{p}
\times \Pi_{p+r} \;, \forall p$ \footnote{Which takes values in a
representation of the Lie group.}. We take $\Phi = (\phi_p ,
\phi_{p+r})$ and $(\a , \b_{r}) \in \Pi_{0} \times \Pi_{r}$, and
the $r$-generalized connection reads as ${\cal A}=(A_{1}, B_{r+1})
\in \Pi_{1} \times \Pi_{r+1} $ and so on. In this case, in view of
(\ref{transdob0}), the symmetry can be written as below:
 \be\label{transdobk} \d
\left(
\begin{array}{c}
  \phi_{p} \\
 \phi_{p+r}
\end{array}\right)=  i \left( \begin{array}{cc}
  \a & \b_r \\
  \b_r & \a
\end{array}\right)
\left(\begin{array}{c}
  \phi_{p} \\
  \phi_{p+r}
\end{array}\right)\; ,
\ee with the $b$-product rule defined in Section 2, where, in the
$b$-term $\sim b\,\b_{r}\phi_{p+r}$, we mean that $r$ indices are
contracted using the metric. Once more, this product leads to a
(non-associative) quasi-group and such that a Lie group is
recovered for $b=0$ or, equivalently, to leading order in $\b$.

We introduce the {\it partial derivative} of a $(0,p)$ tensor
$T_{p}$ as a $(0,p+1)$ tensor given by
\[
T_{p}=T_{\mu_{1}...\mu_{p}}dx^{\mu_{1}}\otimes...\otimes
dx^{\mu_{p}},
\]
as
\[
\partial T_{p}:=\partial_{\mu}T_{\mu_{1}...\mu_{p}}dx^{\mu}\otimes dx^{\mu_{1}
}\otimes...\otimes dx^{\mu_{p}}.
\]
So, we can define the partial derivative of a doublet as the
doublet consisting of the partial derivatives \be
\partial (\phi_p , \phi_{p+r})
\equiv ( (\partial \phi_{p})_{p+1} , (\partial
\phi_{p+r})_{p+r+1}).\ee It is easy to verify that this definition
is consistent with the Leibnitz rule for the product of doublets.

Next, let us give some useful definitions for the associative
case: the tensor product of two doublets of arbitrary orders and
types,
 is the simple generalization of the rule (\ref{defprod}):$
    (A,B)(A',B') = (A \otimes A'  , A \otimes B' + B \otimes A' )$.
We also denote by ${\hat X}$ the totally anti-symmetrized part of
a $(0,p)$ tensor $X$. When applied to doublets, we define it as
$\widehat{(x,y)} \equiv ({\hat x}, {\hat y})$. Furthermore, we can
 define the covariant
derivative of a $p$-doublet (of $r$-order), $\Phi_p =(\phi_p ,
\phi_{p+r+1})$, as a $(p+1)$-doublet: \be D \Phi_p =
\partial \Phi_p - i {\cal A} \Phi_p = \left(
\partial \phi_p- i  A_1 \phi_p \; , \;
\partial \phi_{p+r} - i  A_1 \phi_{p+r} -i B_{r+1} \phi_{p}
\right)\, ,\ee where the connection must be a $1$-doublet ${\cal
A} \equiv ( A_{1} , B_{r+1})$ of order $r$.

 Imposing that $ g D \Phi_{p} = D'
\Phi'_{p} $, and using $ D' =
\partial - i {\cal A}'$ and $\Phi'_{p} =g \Phi_{p}$, we obtain the
transformation law for the connection:\be {\cal A}' = g(\a ,\b)
{\cal A} g(-\a ,-\b) -i( \partial g(\a ,\b)  ) g(-\a ,-\b), \ee
whose infinitesimal expression is $\d {\cal A} =\partial (\a ,\b)
- i[{\cal A},(\a ,\b)]= D (\a ,\b)$\footnote{The canonical
curvature tensor ${\cal F} \in \Pi_{2} \times \Pi_{2+r} $
transforms as ${\cal F}´ = g(\a ,\b){\cal F}g(-\a ,-\b) $.} ;
which reads, in terms of the doublet components: \be \label{A} \d
A =
\partial \a - i [ A, \a ] \; , \ee \be \label{B} \d B = \partial
\b -i [ B, \a ] -i [ A, \b ]\; . \ee

\section{Gauge Fields: BF/Chern-Simons Correspondence\\
 and Yang-Mills Models.}

From now on, we will concentrate on
 separable (associative) groups and doublet representations
  of order $1$, in order to point out some relevant differences
   with previous similar approaches in which $B_{\mu\nu}$ is
    viewed as gauge fields, but without a precise description of
     the underlying symmetry.

So, in terms of tensor components, the curvature tensor ${\cal F}
= ( F_2 , H_3 ) \in \Pi_{2} \times \Pi_{3} $ results as \be
\label{F} F_{\mu \nu} = 2
\partial_{[\mu} A_{\nu]} + i [ A_{\mu} , A_{\nu} ] \ee \be
\label{H} H_{\mu \nu \r} = 2\partial_{[\mu} B_{\nu] \r} + 2i (
B_{[\mu | \r} \, A_{\nu]} - A_{[\mu} B_{\nu]\r} )\; , \ee where
the symbol $|$ before the $\r$ index means that $\r$ is not to be
anti-symmetrized. Since we are considering $(0,p)$ tensors in our
construction and not only $p$-forms, this curvature differs from
the one considered in other approaches where the two-form field is
considered, for us, $H$ {\it is not} totally anti-symmetric but it
contains more components. The Kalb-Ramond gauge field must then be
identified with anti-symmetric part, ${\hat B}\equiv
B_{[\mu\nu]}$.

Next, we may define the {\it topological} Abelian Chern-Simons
Action for the connection ${\cal A}= (A , B)$ as: \be {\cal
S}_{CS} [{\cal A}] \equiv -\frac{k}2 \;\int\; {\cal A} \wedge
{\hat {\cal F}} \equiv -\frac{k}2 \;\int\; [ A \wedge {\hat H} +
 {\hat B} \wedge F ]\, , \label{SC}\ee
where $k$ denotes the inverse of the coupling constant.
 This is a well
defined gauge invariant topological theory which generalizes to a
non-Abelian group as: \be {\cal S}_{CS} [{\cal A}] \equiv -k
\;\int\; \textrm{tr} \; \left( [ A \wedge \partial \wedge {\hat
B} + {\hat B} \wedge \partial \wedge A ] -  i 2 [A \wedge A\wedge
{\hat B}] \right)\, .\label{SCnab}\ee

It is indeed straightforward to check out that ${\cal S}_{CS}$ is
gauge invariant (up to a total derivative) as
expected\footnote{This may be verified by doing first order
variations in $(\a,\b)$ and by using the Bianchi identity (which
is a consequence of Jacobi's identity and the definition of ${\cal
F}$), in the same way that it is usually done for a Chern-Simons
theory in $3d$.}. $B \wedge F $ theories are similar to
 Chern-Simons in three dimensions and they are often formally identified, however,
the actual connection between both never was clearly established
\cite{dob}.
 In the present framework,
by defining $\wedge$ for doublets as the totally anti-symmetrized
tensor product (and taking the integral of the second component),
this can be formulated as
 {\it a genuine} Chern-Simons theory for a doublet connection.

As a result, we observe that a self-interacting $B$-field can only
be obtained for a non-associative gauge symmetry. In fact, as
discussed in Section 2, even in the non associative case (where,
for instance, $b=1$), one could define a gauge theory, since the
infinitesimal algebra would also lead to gauge invariant models.
In this case, one should add a term $ -i[B_{\mu \nu}, \b^{\nu} ] $
in expression (\ref{A}), and $i [ B_{\mu \a} , B_{\nu \b} ]g^{\a
\b} $ in (\ref{F}), however the Chern-Simons theory is not longer
 topological.

In the presence of a spacetime metric, $g$, there exists a natural
map $ A_p \longmapsto A^p \;\,(\Pi_p \to \Pi^{*}_{p}) $ (rising
the indices in ordered way), where $\Pi^{\star}_{p}$ is the {\it
dual} space to $\Pi_{p}$, defined as the set of linear maps, $
<A^p ;... > : \Pi_{p} \to \Re $. This is linearly extended to
doublets through the definition: \be <(A^p , B^{p+1}) ; (C_p ,
D_{p+1})> = < A^p ; C_p > + m_p < B^{p+1} ; D_{p+1} >\, , \ee
where $m_p$ is a real constant which may depend of $p$\footnote{We
may define arbitrarily an internal metric in each two dimensional
doublet space.}. In this way, the operation $<\, ;\,>$ may be
naturally extended to {\it pairs} of doublets, $(A_p , B_{p+1})$.
Therefore, we may define the Yang-Mills Lagrangian for the
generalized connection by: \be {\cal L}_{YM}[( A_\mu , B_{\nu
\sigma})] \equiv -\frac{1}{4\mu^2} \; \, \textrm{tr} <{\cal F} ;
{\cal F} > = -\frac{1}{4\mu^2} \; (\textrm{tr}F^2 \, + m_2
\,\textrm{tr}H^2) , \label{YM}\ee where $\textrm{tr}$ is the trace
in the Lie algebra and $\mu$ is the coupling constant. This model
is gauge invariant {\it only} in the special case
$[A,\b]=0$\footnote{In the separable case ( $G\sim G_{(\a)} \times
G_{(\b)} \, , \;[\a,\b]=0 $) that we are emphasizing here, the
connection may be defined to satisfy this.} , since in this case
the curvature transformations: $( F' , H') = ( e^{i\a} F e^{-i\a}
, e^{i\a} (H - i [F,\b]) e^{-i\a}) = ( e^{i\a} F e^{-i\a} ,
e^{i\a} H e^{-i\a}) = e^{i\a}( F , H )e^{-i\a} $. This is an
remarkable result since, in this formalism, the standard
Lagrangian (\ref{YM}) is gauge invariant in this special case
(which contains the Abelian one) but, {\it a Yang-Mills-type
Lagrangian invariant for a general group symmetry, should require
a functional of higher-order in $F,H$}.

It is remarkable that the theory $S({\cal A}) := {\cal S}_{CS} +
{\cal S}_{YM}$ coincides with the Cremmer-Scherk-Kalb-Ramond model
(rigorously generalized here to non-Abelian groups), which is a
gauge model with massive modes. A crucial ``no go" result in this
type of theories has been presented in ref. \cite{h}. However, it
is interesting to analyze this theory in view of the gauge group
structure clarified here. Since this negative result is based on
the impossibility of closing the BRS algebra, we expect that our
group structure could be crucial in doing that and, thus, in
proving the consistency of this topologically massive model (which
can be an alternative to the Standard Model). Our work will
continue along these lines and the results will be presented in a
forthcoming work \cite{teorkr}.

\section{Coupling with Matter Fields via Minimal Substitution.}

Let us consider the matter free Lagrangian for a doublet of
complex fields $\Phi= (\phi , \phi_\mu)$. Let us denote its
complex conjugate by  ${\bar \Phi}= ({\bar \phi} , {\bar
\phi}_\mu)$ which takes values in a representation of the Lie
group $ G_{(\a)} \times G_{(\b)} $. We can choose a number of
Lagrangians for non-interacting matter \footnote{This means that
it may to interact
 with itself but not with gauge fields.}, \be {\cal L}_{G} = {\cal L}
[{\bar \Phi}\,, \,\Phi \, , \, \partial {\bar \Phi} \, ,
\,\partial \Phi ] , \ee with global gauge invariance.
 Thus,
if we consider the group parameters as local, this Lagrangian
becomes locally gauge invariant if we perform ``minimal
substitution", {\it i.e.}, if we replace partial derivatives by
covariant ones. Doing so, we obtain the full Lagrangian density
which contains {\it minimal} interactions involving one and
two-form gauge field:
 \be {\cal L}_{L} [{\bar \Phi}, \Phi , \partial {\bar \Phi}  , \partial \Phi , A_\mu
, B_{\nu \r}] = {\cal L}[{\bar \Phi}\,, \,\Phi \, , \, {\bar D}
{\bar \Phi} \, , \, D \Phi ],\ee
 where \be D \Phi = [\partial - i {\cal
A}] \Phi = \left( \partial_\mu \phi - i  A_\mu \phi \; , \;
\partial_\mu \phi_\nu - i  A_\mu \phi_\nu -i B_{\mu \nu} \phi
\right)\, ,\ee and\be {\bar D} {\bar \Phi} = [\partial + i {\cal
A}] {\bar \Phi} = \left( \partial_\mu {\bar \phi} + i  {\bar \phi}
A_\mu \; , \;
\partial_\mu {\bar \phi}_\nu + i {\bar \phi}_\nu A_\mu  +  i {\bar \phi}  B_{\mu \nu}
\right)\, .\ee

 The simplest globally invariant Lagrangian, which is second order and
  quadratic in the fields, involves
only the first component (scalar) of the doublet: \be \label{KG}
{\cal L}_{scalar} = \frac12  \partial^\mu {\bar \phi} \partial_\mu
\phi - \frac{M^2}{2}  {\bar \phi} \phi .  \ee However, it is not
very interesting because, when we replace the partial
 by the covariant derivatives,
the only interaction that appears is with the 1-form gauge field
$A_\mu$,
 as expected for a complex scalar field. So, in order to
get matter interacting with $B_{\mu \nu}$, one must to consider
higger order
 Lagrangians (in the matter fields).
The combinations \be\Psi_0 = {\bar \Phi} \Phi \; , \; \Psi_1 =
{\bar \Phi} \,
 \partial_\mu \Phi \; ,
\;  {\bar \Psi_1 } = (\partial_\mu {\bar \Phi})\, \Phi \; ,\;
\Psi_2 =
 \partial_\nu {\bar \Phi} \partial_\mu \Phi , \dots \label{invariants}\ee and
  so on,
are invariant doublets under global gauge transformation and we
can form Lagrangian
 densities by
constructing scalars with them. For instance, using the previous
definition of the
 internal product, we may write down
free Lagrangian densities in addition to (\ref{KG}):
\be\label{freemat} {\cal L}_G = \frac14 \left(c_0 < \Psi_0 \, ; \,
\Psi_0 > +
  c_1 < {\bar \Psi}_1 \, ; \, \Psi_1 > +  c_2 < \Psi_2 \, ; \, \Psi_2 > + \dots \right),\ee
where $c_{1,2,3,...}$ are constant coefficients.

Now, using the standard rule of the Noether's theorem for the
symmetry described here,
 we are able to find new conservation laws associated to the interaction with both,
  rank 1 and 2 gauge fields.
 The corresponding Noether's currents are:
 \be\label{j1}
 J^\mu = i \phi \frac{\partial {\cal L}}{\d \partial_\mu \phi } - i
  {\bar \phi} \frac{\partial {\cal L}}{\d \partial_\mu  {\bar \phi} }
 + i \phi_\nu \frac{\partial {\cal L}}{\d \partial_\mu \phi_\nu }- i
  {\bar \phi}_\nu \frac{\partial {\cal L}}{\d \partial_\mu {\bar \phi}_\nu } \, ,
 \ee
and remarkably
 \be\label{j2}
 J^{\mu \nu} = i \phi \frac{\partial {\cal L}}{\d \partial_\mu \phi_\nu }- i
  {\bar \phi} \frac{\partial {\cal L}}{\d \partial_\mu {\bar \phi}_\nu } \, ,
 \ee
which, by virtue of the (free) equations of motion, satisfy
$(\partial_\mu
 {\cal J}^\mu \, ; \, \partial_\mu {\cal J}^{\mu \nu}) = 0$.

As mentioned, we localize this symmetry by substituting $\partial$
by $D$ in
 the doublets $\,{\bar \Psi}_1 \,, \,\Psi_1 \,, \,\Psi_2 \, \dots$,
which, for construction, become {\it locally invariant} objects;
so finally,
 replacing them into (\ref{freemat}), the proper locally gauge invariant
  Lagrangian interacting with the gauge fields $(A ,B)$
is canonically obtained.

As an example consider the simplest model, where an {\it Abelian}
$B$-interaction arise from a gauge principle. Let us take $c_{i
\geq 2}\equiv 0$ in eq. (\ref{freemat}), which is quartic in the
matter fields; so, we consider: \be \label{KGfreemat} {\cal
L}_{Matter} \equiv  \frac 14 \left(c_0 < \Psi_0 \, ; \, \Psi_0 > +
  c_1 < {\bar \Psi}_1 \, ; \, \Psi_1 > \right), \ee
where explicitly: \be \label{mat0} < \Psi_0 \, ; \, \Psi_0 > =
\phi^2 {\bar \phi}^2 +
 (\phi {\bar \phi}_\mu + {\bar \phi} \phi_\mu )(\phi^\mu {\bar \phi} + {\bar \phi}^\mu \phi )\ee
and \be\label{mat1} < {\bar \Psi}_1 \, ; \, \Psi_1 > = \phi {\bar
\phi} \partial_\mu \phi
 \partial^\mu {\bar \phi} + (\phi \partial^\nu {\bar \phi}^\mu +
 \phi^\mu \partial^\nu {\bar \phi})( {\bar \phi} \partial_\nu \phi_\mu  +
 {\bar \phi}_\mu \partial_\nu \phi ).\ee This may be viewed as a
 Sigma model whith a non-trivial metric on the fields manifold.

In components, the gauge transformations read as
\bea\label{expsim}
\phi' = e^{i\a}\phi ~~~;~~~\phi'_\mu = e^{i\a} (\phi_\mu + i\b_\mu \phi)\nonumber\\
{\bar \phi}' = {\bar \phi} e^{- i\a} ~~~;~~~{\bar \phi}'_\mu =
({\bar \phi}_\mu - i\b_\mu {\bar \phi})e^{-i\a} , \eea  which are
global symmetries of ${\cal L}_{Matter} $ as can be easily
verified. In order to preserve this symmetry when the parameters
are considered functions of the spacetime point, one must to
replace $\partial_\mu$ by covariant derivatives in expression
(\ref{mat1}), which according to our definitions reads:
\bea\label{mat1} < {\bar \Psi}_1 \, ; \, \Psi_1
>_{local} = \phi {\bar \phi}(\partial_\mu - i A_\mu )\phi \;\;
 (\partial^\mu + i A^\mu ) {\bar \phi} +
( \phi (\partial_\nu + i A_\nu ) {\bar \phi}_\mu +\nonumber\\
  \phi_\nu (\partial_\mu + i A_\mu ){\bar \phi})
({\bar \phi} \; (\partial^\nu - i A^\nu )
  \phi^\mu  +
   {\bar \phi}^\nu (\partial^\mu - i A^\mu )\phi )
- i( \phi (\partial^\mu + i A^\mu ) {\bar \phi}^\nu +
  \phi^\mu (\partial^\nu + i A^\nu ){\bar \phi})
B_{\mu \nu} \phi
  {\bar \phi}+\nonumber\\
i ({\bar \phi} \; (\partial^\mu - i A^\mu )
  \phi^\nu  +
   {\bar \phi}^\mu (\partial^\nu - i A^\nu )\phi )
  B_{\mu \nu} \phi {\bar \phi} +
  B_{\mu \nu}B^{\mu \nu} \phi^2 {\bar \phi}^2
.\eea This manifestly describes the minimal interaction of the
$B$-field
 with the matter. So, from expressions (\ref {j1}),
we have a standard current density \bea J^{\nu}&=& \frac{i}{4} c_1
(\phi^2 {\bar \phi}
\partial^\nu {\bar \phi} - \phi {\bar \phi}^2
\partial^\nu \phi ) + \nonumber\\
&+&\frac{i}{4} c_1 [ (\phi {\bar \phi}_\mu + \phi_\mu {\bar \phi})
(\phi
\partial^\nu {\bar \phi}^\mu + \phi^\mu \partial^\nu {\bar \phi})
- (\phi {\bar \phi}_\mu + \phi_\mu {\bar \phi}) (  {\bar \phi}
\partial^\nu \phi^\mu + {\bar \phi}^\mu \partial^\nu \phi)],
\eea where we have ignored the contribution of the additional
Klein-Gordon Lagrangian (\ref{KG}). Furthermore, according to
(\ref {j1}), we also have the tensorial current: \be J^{\mu \nu}=
\frac{i}{4} c_1 \phi {\bar \phi}[\phi
\partial^\mu {\bar \phi}^\nu -
 {\bar \phi} \partial^\mu \phi^\nu  + \phi^\mu \partial^\nu {\bar \phi} -
 {\bar \phi}^\mu \partial^\nu \phi]\, .
\ee This result is new. It reveals the conservation of a charge
that has a vector index and arises from a gauge symmetry, this is
not however the momentum generator, as the symmetry is not a
translation. Indeed, tensor-like charges may appear in higher
dimensions whenever a supersymmetry algebra is settled with
central charges \cite{luk}.

\section{Concluding Remarks.}

In this paper we have found many answers to old questions related
to the $B\wedge F$ field theories (where $B$ is a KR field). We
cleared the group structure underlying these models and
constructed the KR-field through a standard connection even for
non-Abelian Lie groups, thus setting the formalism to decide
definitively if well defined topologically massive models are
possible or not in four space time dimensions. Finally, for the
first time also, we built a theory where there is minimal
interaction with a (gauge) tensor field and we found a conserved
current associated with its gauge character (via Noether's
theorem). Many open possibilities on both, mathematics and
physics, have been briefly pointed out in this article, which will
be developed properly elsewhere.

Among several possible applications of this framework, we stress
that perhaps it could be helpful in formulating gravitation as a
genuine topological ($B\wedge F$) theory which nowadays is an
strong research line; and
 in a mathematical context,
 it could provide new insights
in order to find topological invariants in four or more dimensions
and novel
 realizations of non-Associative algebras \cite{asoc}.

{\bf Acknowledgements}: The author is indebted to A. L. M. A.
Nogueira, J. A. Helay\"{e}l-Neto and S. A. Dias for invaluable
suggestions, comments and criticisms. Special thanks are due to
Prof. G. Thompson for helpful criticisms and important
suggestions,
 and to Prof. A. Lahiri
 for stimulating comments on this research line. The
author is supported by CLAF/CNPq.

\end{document}